\documentclass[apjl,iop]{emulateapj}

\newcommand{\vB}{{\bf B}}
\newcommand{\vJ}{{\bf J}}
\newcommand{\vv}{{\bf v}}
\newcommand{\del}{\nabla}
\newcommand{\Section}{{\S}}
\newcommand{\um}{{\,\mu\rm m}}
\newcommand{\cm}{{\rm\,cm}}

\newcommand{\au}{{\rm\,AU}}

\newcommand{\yr}{{\rm\,yr}}

\newcommand{\gm}{{\rm\,g}}
\newcommand{\gram}{{\gm}}

\newcommand{\cms}{{\rm\,cm\,s^{-1}}}

\newcommand{\msun}{{M_\odot}}

\newcommand{\K}{{\rm\,K}}
\newcommand{\kelvin}{{\K}}
\newcommand{\second}{{\rm\,s}}
\newcommand{\muG}{{\,\mu\rm G}}
\newcommand{\etaunit}{{\cm^2\second^{-1}}}
\newcommand{\Hydrogen}{{{\rm H}}}
\newcommand{\Htwo}{{\Hydrogen_2}}
\newcommand{\mH}{{m_\Hydrogen}}

\newcommand{\mHtwo}{{m_\Htwo}}
\newcommand{\nHtwo}{{n_\Htwo}}
\newcommand{\N}{{{\rm N}}}

\begin{document}

\slugcomment{Submitted to ApJ 2009 October 21; accepted 2010 April 2}
\journalinfo{{\rm Also available in} The Astrophysical Journal, {716, 1541--1550}}

\title{Disk Formation Enabled by Enhanced Resistivity}

\author{Ruben Krasnopolsky\altaffilmark{1,2}, Zhi-Yun Li\altaffilmark{3,2}, Hsien Shang\altaffilmark{1,2}}
\altaffiltext{1}{Academia Sinica, Institute of Astronomy and Astrophysics, Taipei, Taiwan}
\altaffiltext{2}{Academia Sinica, Theoretical Institute for Advanced Research in Astrophysics, Taipei, Taiwan}
\altaffiltext{3}{University of Virginia, Astronomy Department, Charlottesville, USA}

\shortauthors{{\sc Krasnopolsky, Li, and Shang}}
\shorttitle{{\sc Disk Formation Enabled by Enhanced Resistivity}}

\begin{abstract}
Disk formation in magnetized cloud cores is hindered by magnetic braking.
Previous work has shown that for realistic levels of core magnetization,
the magnetic field suppresses the formation of rotationally supported
disks during the protostellar mass accretion phase of low-mass star
formation both in the ideal MHD limit and in the presence of ambipolar
diffusion for typical rates of cosmic ray ionization. Additional effects,
such as ohmic dissipation, the Hall effect, and protostellar outflow,
are needed to weaken the magnetic braking and enable the formation of
persistent, rotationally supported, protostellar disks. In this paper,
we first demonstrate that the classic microscopic resistivity is not
large enough to enable disk formation by itself. We then experiment
with a set of enhanced values for the resistivity in the range
$\eta=10^{17}$--$10^{22}\etaunit$. We find that a value of
order $10^{19}\etaunit$ is needed to enable the formation
of a $10^2\au$-scale Keplerian disk; the value depends somewhat on
the degree of core magnetization. The required resistivity is a
few orders of magnitude larger than the classic microscopic values.
Whether it can be achieved naturally during protostellar collapse
remains to be determined.
\end{abstract}
\keywords{accretion, accretion disks --- ISM: clouds --- ISM: magnetic fields --- magnetohydrodynamics (MHD)}

\section{Introduction}

\subsection{Magnetic Braking and Disk Formation}

Circumstellar disks play a central role in both star and planet
formation. The disks are thought to be the conduit through which
Sun-like stars assemble the bulk of their mass \citep{sal87}.
They are also the birthplace of planets. Despite years of active
research (see reviews by \citealp{b95} for
early work), we still do not have answers to some of the most
basic questions, such as ``when and how does a disk appear in
the process of star formation?''

Observationally, it is difficult to separate out the protostellar
disk from the surrounding envelope during the early, embedded
phase of star formation with the current generation of interferometers
(e.g., \citealp{j07}), although the situation will improve
drastically in a few years, when ALMA comes online \citep{dj08}.
\citet{ecdd09} inferred the existence of a large (hundreds of AU
in radius), massive (more than $0.1\msun$) disk around the
protostar Serpens FIRS 1, based mostly on CARMA dust continuum
observations. Without a detailed knowledge of the central stellar
mass and gas kinematics on small (arcsec) scales, it is difficult
to determine whether the inferred structure is a rotationally
supported disk, a magnetically induced ``pseudodisk,'' or simply
the inner part of an envelope that is dynamically more complicated
than the \citeauthor{u76}'s (\citeyear{u76}) rotating, collapsing sphere
adopted by \citet{ecdd09} for the protostellar envelope.
Theoretically, disk formation in dense cores of molecular
clouds magnetized to
the observed level (corresponding to a mean dimensionless
mass-to-flux ratio $\lambda \sim$
a few; \citealp{tc08})
is controlled by magnetic braking, which has been difficult to
quantify until recently.

Detailed calculations have now shown that the magnetic braking is
apparently so efficient as to inhibit the formation of rotationally
supported disks {\it during the main protostellar accretion phase}
for a moderate level of core magnetization, as long as the
magnetic field is frozen in the matter (i.e., the ideal MHD limit;
\citealp{als03}; \citealp{glsa06}; \citealp{pb07};
\citealp{hf08}; \citealp{ml08};  see, however,
\citealp{hc09} for a different view,  and also
\citealp{mmth05} and \citealp{bp06} for the evolution
of rotating, magnetized cores prior to the main accretion phase).
In order for a rotationally supported disk to exist around a
rapidly accreting protostar, the magnetic braking must be weakened
one way or another. The most obvious possibility is through
non-ideal MHD effects, which are expected to be important
in lightly ionized star-forming cores. However, we have recently
shown that ambipolar diffusion, the best studied non-ideal MHD
effect in star formation, does not sufficiently weaken the
braking to allow rotationally supported disks to form for
realistic levels of core magnetization and ionization; in
some cases, the magnetic braking is even enhanced (\citealp{ml09},
\citealp{kk02}; see \citealp{bm95}, \citealp{hw04}
and \citealp{dp09}, who considered
the effects of ambipolar diffusion on magnetic braking and
core fragmentation during the earlier phase of prestellar core
evolution). This motivates us to examine whether two of the remaining
non-ideal MHD effects, ohmic dissipation and the Hall
effect, can weaken the coupling between the magnetic field
and the bulk neutral matter (and thus the strength of
magnetic braking) enough to enable the formation of
rotationally supported disks. This paper concentrates on
the ohmic dissipation.

\vskip 1ex

\subsection{Electrical Resistivity of Cloud Cores}

The resistivity $\eta$ is related to the electrical conductivity
$\sigma$ through
\begin{equation}
\eta = \frac{c^2}{4\pi\sigma}\ .
\label{resistivity}
\end{equation}
For lightly ionized molecular gas in dense cores, the electric
conductivity is usually dominated by the contribution from
electrons, although the grain contribution can dominate at
high densities when large amounts of small grains are present
(e.g., \citealp{wn99}). To obtain a rough estimate for the
conductivity, we will concentrate on the contribution from
the electrons, which is given by
\begin{equation}
\sigma_e = \frac{ n_e e^2 \mHtwo}{m_e \rho \langle\sigma v\rangle_{e-\Htwo}}
= 9.0 \times 10^{16} x_e\ (\second^{-1})\ ,
\label{conductivity}
\end{equation}
where $\rho = 2.8 \mH \nHtwo$ and $x_e=n_e/\nHtwo$.
We have adopted a value $\langle\sigma v \rangle_{e-\Htwo}
=2\times 10^{-9}\cm^3\second^{-1}$ for a $10\kelvin$ gas from
\citet{pg08}.  This value is somewhat smaller
than that used in \citet{drd83} and \citet{wn99}, which
is $2.6\times 10^{-9}\cm^3\second^{-1}$, but somewhat
larger than that of \citet{smun00}, which is $1.3\times 10^{-9}\cm^3\second^{-1}$.
It increases slowly with temperature, roughly as $T^{1/2}$,
yielding a lower
resistivity at a higher temperature. Since our calculation is assumed to be
isothermal, the temperature dependence is not used.

Substituting equation (\ref{conductivity}) into equation
(\ref{resistivity}) yields
\begin{equation}
\eta = \frac{7.9\times 10^2}{ x_e}\ (\etaunit)\ .
\end{equation}
The fractional ionization $x_e$ can be computed from detailed
chemical networks. At the relatively high densities that
are relevant for disk formation, it depends on the grain
size distribution. For the standard \citet{mrn77}
distribution, \citet{nnu02}
computed the ionization fraction over a large
range of number density (their Figure 1). Over the
critical range of density $\sim 10^8$--$10^{12}\cm^{-3}$
where the transition between the rapidly infalling
pseudodisk\footnote{By ``pseudodisk'' we mean in this paper a structure
formed by the magnetically induced flattening of infalling material
due to anisotropic magnetic forces, as in \citet{gs93}.}
and a potential Keplerian disk is expected to occur,
the fraction can be fitted roughly by
\begin{equation}
x_e \approx 10^{-14}
\left(\frac{10^{10}\cm^{-3}}{\nHtwo}\right)\ ,
\label{ionization}
\end{equation}
which corresponds to a resistivity of
\begin{equation}
\eta = 1.7\times 10^{17}
 \left(\frac{\rho}{10^{-13}\gram\cm^{-3}}\right)\ (\etaunit)\ .
\label{standard}
\end{equation}
The fit of fractional ionization deviates substantially
from the computed values at low densities. However, the
deviation is expected to be inconsequential because the
low density envelope is sufficiently ionized to be close
to the ideal MHD limit (ignoring ambipolar diffusion,
which is investigated separately elsewhere, see e.g.,
\citealp{kk02} and \citealp{ml09})
according to either equation (\ref{ionization}) or the
computed values. If only relatively large grains of
size (of order $0.1\um$) exist, the ionization fraction
would be higher (by as much as three orders of magnitude,
see Figure 1 of \citealp{un90}, or compare
Figures 1 and 3 of \citealp{wn99}), making the resistivity
smaller. Equation (\ref{standard}) should therefore be
viewed as a rough indication of the plausible upper range
of the ``classic'' (as opposed to ``anomalous'') resistivity,
especially in view of the fact that the actual temperature
in the high density region of disk formation is expected to
be higher than the $10\kelvin$ adopted in our calculation, which would
lower the resistivity.
It may be enhanced, however, through anomalous processes
(e.g., \citealp{nh85}).

The rest of the paper is organized as follows. In \Section{}2,
we describe the setup of the disk formation problem,
focusing on the protostellar mass accretion phase,
when a central object has formed. For illustrative
purposes, we will concentrate on a particular set of
core parameters for which a $10^2\au$-scale disk
is formed at a representative time $t=10^{12}\second$
(or about $3\times 10^4\yr$) in the absence of
magnetic braking, and a magnetic field that is
strong enough to suppress the formation of a
rotationally supported disk in the ideal MHD limit.
We consider in \Section{}3 whether the illustrative classic
resistivity given in equation (\ref{standard}) can
weaken the magnetic braking enough to enable disk formation,
and find that the answer is no. In \Section{}4, we explore
the question of how large the resistivity must be
in order for the $10^2\au$-scale rotationally supported
disk to reappear in our model problem. It turns out
to be of order $10^{19}\etaunit$, a few
orders of magnitude larger than the classic value expected
on the $10^2\au$-scale, as anticipated by \citet{sglc06}
based on the concept of the so-called Ohm sphere,
although our value is one to two orders of
magnitudes smaller than theirs, depending on the field strength.
The numerical results and their astro\-physical
implications are discussed in the last section, \Section{}6.

\section{Problem Setup}

Our primary goal is to illustrate the effects of ohmic
dissipation on magnetic braking and disk formation.
For this purpose, it is desirable to have as small
a numerical magnetic diffusivity as possible, which
demands a high spatial resolution. High resolution,
however, puts a stringent limit on the computation
time step $dt$, which is proportional to the square
of the (smallest) grid size for the explicit method
that we use to treat the ohmic dissipation. Although
subcycling can be used to alleviate the time step
problem substantially, a large number of cycles are still
needed to reach a reasonable protostellar accretion
time, say, $10^{12}\second$, making the treatment of
self-gravity difficult.  For this initial exploration
of the effects of ohmic dissipation, we have decided
to set up a simplified disk formation problem, with
a rotating, non-self-gravitating envelope falling
onto a central object of fixed mass $M_\ast$.  It will
turn out that a large ``anomalous'' resistivity is
needed to save the disk, an elaborate disk formation
model is probably not warranted at this early stage
of the investigation.

Specifically, we solve the equations of resistive MHD,
\begin{equation}
\frac{\partial\rho}{\partial t}
+\del\cdot(\rho\vv)=0\ ,
\end{equation}
\begin{equation}
\frac{\partial\vv}{\partial t}+
(\vv\cdot\del)\vv=-\frac{\del p}{\rho}+\frac{\vJ\times\vB}{\rho c} -\del\Phi_g\ ,
\end{equation}
\begin{equation}
4\pi \vJ = c\del\times\vB\ ,
\end{equation}
\begin{equation}
\frac{\partial\vB}{\partial t}=\del\times(\vv\times\vB-\eta\del\times\vB)
\end{equation}
under the assumption of axisymmetry, and adopt a spherical
polar coordinate system $(r, \theta, \phi)$, with a
stellar object of $M_\ast=0.5\msun$ at the origin such that
the gravitational potential is $\Phi_g= - GM_\ast/r$,
neglecting the self-gravity of the gas. Self-gravity
will be included in a follow-up work, where we will evolve the
magnetized core from its formation to collapse in a self-consistent
manner. For this initial study, we aim to capture the essence of the
problem of magnetic suppression of disk formation in the simplest
possible way, by adopting uniform distributions for the initial
gas density and magnetic field, and to study the effects of a new
ingredient, resistivity, in this simplest problem.

At time $t=0$, we fill the computation domain between
$r_i=1.5\times 10^{14}\cm$ and
$r_o=1.5\times 10^{17}\cm$ with a uniform density
$\rho_0=1.4\times 10^{-19}\gram\cm^{-3}$ (corresponding
to $n(\Htwo)=3.0\times 10^4\cm^{-3}$), so that
the total envelope mass is $1\msun$. For simplicity,
we assume that the gas stays isothermal, with an
isothermal sound speed $a=(p/\rho)^{1/2}=2\times 10^4\cm\second^{-1}$
(corresponding to a temperature of about $10\kelvin$).
For the initial rotation, we adopt the following prescription:
\begin{equation}
v_\phi = v_{\phi,0} \tanh(\varpi/\varpi_c)
\end{equation}
where we choose $v_{\phi,0}=2\times 10^4\cm\second^{-1}$,
$\varpi$ is the cylindrical radius, and
$\varpi_c=3\times 10^{15}\cm$.
The softening of the rotational profile inside the cylindrical radius
$\varpi_c$ is to prevent the angular speed from becoming singular near
the rotation axis.
Since the inner radius of
our computational domain is $10\au$, we are concerned
with the formation or suppression of relatively
large disks (of tens of AU or more) only.
Our initial rotation is relatively fast: \citet{gbfm93}
estimated observationally
that the ratio of rotational to gravitational energies for dense
$\N\Hydrogen_3$ cores is typically $\sim 0.02$, which would imply $v_\phi \sim
3\times 10^3\cm\second^{-1}$ for our setup. Our choice
of a faster rotational speed is conservative in that it is harder
to remove a larger initial angular momentum through magnetic
braking (see also \citealp{ml08}).

The calculations
are done with a new version of Zeus, dubbed ``ZeusTW''.
ZeusTW, written in idiomatic Fortran95,
is based on the ``Zeus36'' code \citep{klb99,klb03},
itself derived from Zeus3D (LCA version 3.4.2: \citealp{cnf94});
Zeus36 is parallel by domain decomposition, and utilizes dynamic memory for
its field and grid arrays.  A small memory pool of temporary arrays
replaces the worker arrays of Zeus3D,
increasing programming flexibility at essentially zero runtime cost.
ZeusTW adds to Zeus36 the ability to solve many non-ideal MHD problems in an explicit form,
covering the ohmic, Hall, and ambipolar diffusion terms.
To treat the ohmic term relevant for this paper, we used a
resistivity algorithm based on \citet{fsh00},
which includes subcycling.  We tested the code by diffusing
an initial Gaussian profile in Cartesian geometry, in one, two, and
three dimensions.  We also tested slightly different forms of the algorithm
(such as changing the operator splitting by calculating the current
density $\vJ$ before or after the magnetic field $\vB$
is updated), and different subcycling prescriptions (such as
varying the maximum number of ohmic subcycles from 1 to 100).
The code passed all of these tests.

\begin{figure}
\includegraphics[width=\columnwidth]{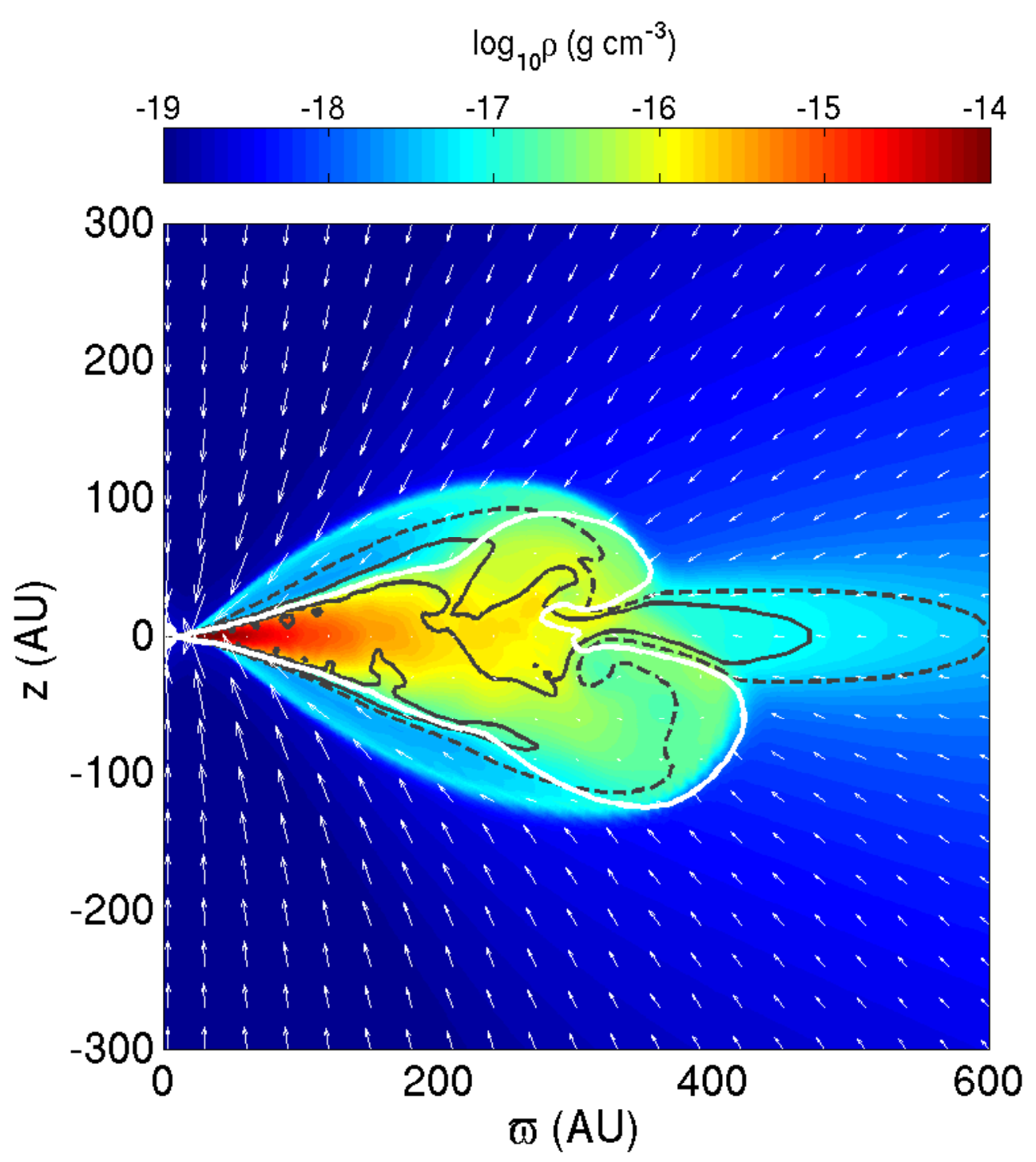}
\caption{Snapshot of the rotationally supported disk formed
in the absence of magnetic braking at a representative time
$t=10^{12}\second$.  Logarithm of density (colormap);
radial infall sonic transition
$v_r=-a=-2\times 10^4\cms$ (white line);
poloidal velocity field (arrows, scale can be estimated from the radial infall component at the white line);
and level of Keplerian support $|v_\phi/v_K|$
(dark gray dashes at 90\% and solid line at 100\%),
where $v_K=(G M_\ast/\varpi)^{1/2}$.}
\label{hydro}
\end{figure}

\subsection{Disk Formation in the Non-Magnetic Case}

We adopt the standard hydro outflow boundary conditions at both the inner
and outer (radial) boundaries of the computation domain
(these boundary conditions impose continuity of material outflowing the grid,
helping to reduce artificial reflections of waves at boundaries).
We use $400$ non-uniformly spaced grid points
in the radial direction, corresponding to a smallest
radial grid size $dr_{\rm min}=3\times 10^{12}\cm$
(or $0.2\au$) and a ratio of adjacent zone sizes of $1.017$.
We use 180 uniform zones in
the $\theta$ direction, corresponding to a minimum
polar grid size of $2.6\times 10^{12}\cm$, comparable
to the smallest radial grid size.
With this setup, we obtain a rotationally supported
disk of radius $r_d\approx 400\au$ and mass
$m_d\approx 0.01\msun$ at a time $t=10^{12}\second$
in the absence of magnetic braking; the time is
comparable to the duration of the deeply embedded Class 0 phase
(\citealp{awb00}, see, however, \citealp{e09}
who concluded that the duration is longer). A snapshot
of the disk is shown in Figure \ref{hydro}.
The hydrodynamic disk is resolved in the radial direction by
more than 200 zones. We have experimented with coarser grids
(e.g., a total of $200$ radial zones and $90$ angular zones),
finding qualitatively similar results, which indicates that the
resolution employed is sufficient.  An example of the lower
resolution runs is shown in \Section{}\ref{sec:er} below.

The disk size at $t=10^{12}\second$ can be understood roughly as
follows. We first estimate the radius of the infall
region $r$ at time $t$, and the specific angular momentum $\ell$
of the material at that radius.
Assuming that pressure gradients are negligible compared to the
central gravity, we have the infall speed at any given radius to be near the free fall value
$v(r) = (2 G M_\ast / r )^{1/2}$. The infall radius is then roughly $t \sim r/v \sim r^{3/2} /(2 G M_\ast)^{1/2}$,
which yields $r \sim 5\times 10^{16}\cm$ at $t=10^{12}\second$, much larger than $\varpi_c$
(the characteristic radius for rotation initial profile change). Therefore, the equatorial material
at radius $r$ has a specific angular momentum of $\ell=rv=r \times 2\times 10^4\cms$.  Conservation of angular
momentum around a star of mass $0.5\msun$ tells us that the specific angular momentum
corresponds to a centrifugal radius of $\sim 1000\au$, which is about a
factor of 2 larger
than our numerically obtained disk radius $\sim 400\au$. We consider the
agreement satisfactory, given the crudeness of the infall estimate, and
the fact that most material on the surface of the infall sphere at radius
$r$ has a specific angular momentum $\ell$ less than the above estimate
(valid only for the equatorial region), and the centrifugal radius is
sensitive to $\ell$ (as $\ell^2$).

\begin{figure}
\includegraphics[width=\columnwidth]{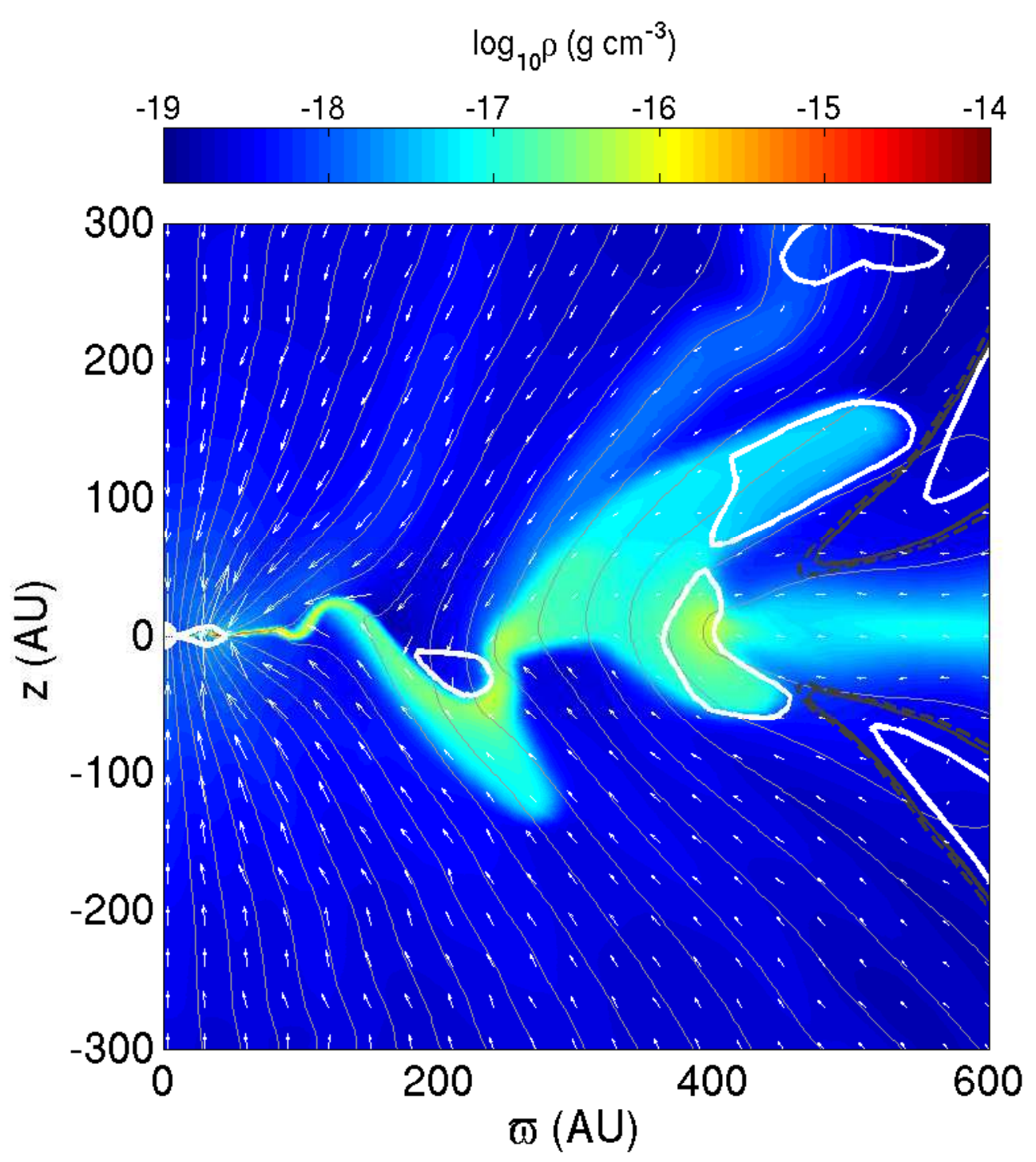}
\caption{Same as Figure \ref{hydro} but for the ideal MHD model. The
field lines are drawn as thin solid lines
(20 contour lines of magnetic flux, with flux levels
quadratically spaced from zero at the axis up to the flux value
at the outer equator point of each magnetized figure).
The field component $B_z$ is $470\muG$ at the center of the figure.
No rotationally supported structure forms in this simulation, although the flow
structure is strongly affected by numerical reconnection
of field lines.}
\label{IdealMHD}
\end{figure}

\subsection{Suppression of Disk Formation in the Ideal MHD Limit}

As mentioned in \Section{}1, previous studies have shown
that a moderate magnetic field corresponding to a dimensionless
mass-to-flux ratio of a few can completely suppress the
formation of rotationally supported disks in the ideal MHD
limit (at least when the rotation and magnetic axes are aligned;
see \citealp{hc09}).
Even though the mass-to-flux ratio is ill-defined in our
current model setup that does not include the self-gravity of
the envelope gas, we find that the disk formation can still be
suppressed if the magnetic field is strong enough. An example
is shown in Figure \ref{IdealMHD}, where the initially uniform
mass distribution in the computational domain of the hydro
model shown in Figure \ref{hydro} is threaded with a uniform
magnetic field $B_0=\sqrt{4\pi}\times10\muG\approx 35\muG$ along
the rotation axis. The field strength is in the range inferred
by \citet{tc08} for a sample of dark cloud cores,
after correcting for projection effects.
For comparison, we also considered a weaker field of
$B_0=\sqrt{4\pi}\times 3\muG \approx 10.6\muG$ and
obtained qualitatively similar results.  For MHD calculations,
we adopt the standard outflow boundary condition
for the magnetic field at the outer radial boundary, and a
torque-free (i.e., $B_\phi=0$) outflow boundary condition at
the inner radial boundary, as in \citet{ml08}.

The flow structure is strongly
affected by numerical reconnection of the field lines. The
reconnection is an unavoidable consequence of the dragging
of the field lines into a highly pinched, split-monopole
type configuration by gravitational collapse in the ideal
MHD limit \citep{glsa06}. The build up of magnetic
flux in the central region as a result of continuous mass
accretion forces the oppositely directed field lines
above and below the equator closer and closer together,
eventually triggering (numerical) reconnection.
It is present in other ideal MHD calculations of
magnetized core collapse, such as \citet{ml08}.
Because of the (unavoidable) numerical artifacts, it is
difficult to quantify the exact strength of the magnetic
braking. Nevertheless, it is clear from Figure \ref{IdealMHD}
that a coherent disk does not exist; it is replaced by
a set of dense blobs, which are not
rotationally supported (and may break up asymmetrically
in 3D). The prominent dense blobs and sheets
at radii $400\au$ and smaller are all rotating
well below their local Keplerian speed (especially the
innermost ones), as can be seen from the thick dashed
and solid dark gray lines at radii greater than $\varpi
\sim 400\au$ in Figure \ref{IdealMHD}, which
mark the location of $|v_\phi|=0.9\,v_K$ and $v_K$, respectively
($v_K$ is the local Keplerian speed). The blobs and sheets
are supported to a large extent by the
magnetic tension force from the pinched field lines.

\section{Inability for Classic Resistivity to Enable Disk
Formation}

We now address the question of whether the illustrative
classic resistivity given in Equation (\ref{standard})
can weaken the magnetic braking enough to enable the
formation of rotationally supported disks or not. For
this purpose, we have implemented into the MHD code ZeusTW
an explicit treatment of resistivity, following the work
of \citet{fsh00} and \citeauthor{ms97} (\citeyear{ms97}; see also \citealp{fc02}).
In these treatments, the resistive term of the induction equation is included through operator splitting,
respecting the constrained transport condition that keeps the field divergence null to machine round-off.
Numerical stability of this term limits the timestep $\Delta t_\Omega$ to a value $\propto (\Delta x)^2$,
a quadratic requirement that can be much more stringent than
the time $\Delta t_{\mathrm{IMHD}} \propto (\Delta x)^1$
required for treating the hydrodynamics and the ideal MHD terms.
Fortunately, the resistive term is computationally inexpensive compared to the
rest of the induction equation, and so it can be efficiently subcycled.
We have checked that subcycling speeds up
the calculations typically by a substantial factor and does
not change the results significantly.
We have also tried small changes in the operator splitting scheme, such as altering
the order of the ideal MHD and the resistivity terms, or adding both terms together.
These changes had no substantial effect on the results,
and therefore we chose the scheme that was most convenient for efficient subcycling,
which is one identical to that used in \citet{fsh00}.

\begin{figure}[t]
\includegraphics[width=\columnwidth]{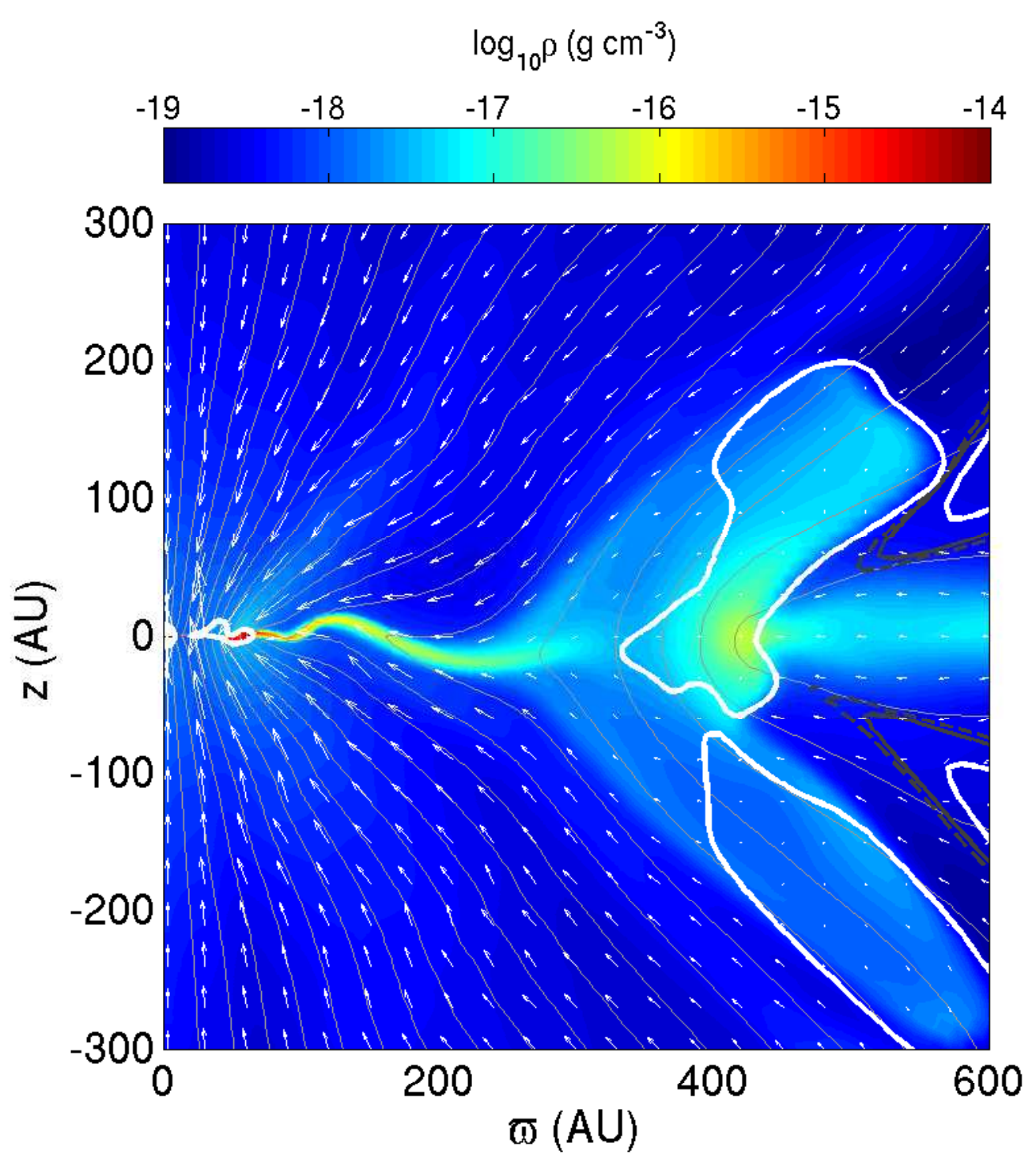}
\caption{Same as Figures \ref{hydro} and \ref{IdealMHD} but for
the MHD model with the classic resistivity given in
Equation (\ref{standard}). A dense, flattened structure exists
near the equator, but it is a highly dynamic magnetically
produced pseudodisk rather than a rotationally supported
disk. The flow structure is affected by
numerical reconnection of field lines.}
\label{classic}
\end{figure}

\begin{figure}
\includegraphics[width=\columnwidth]{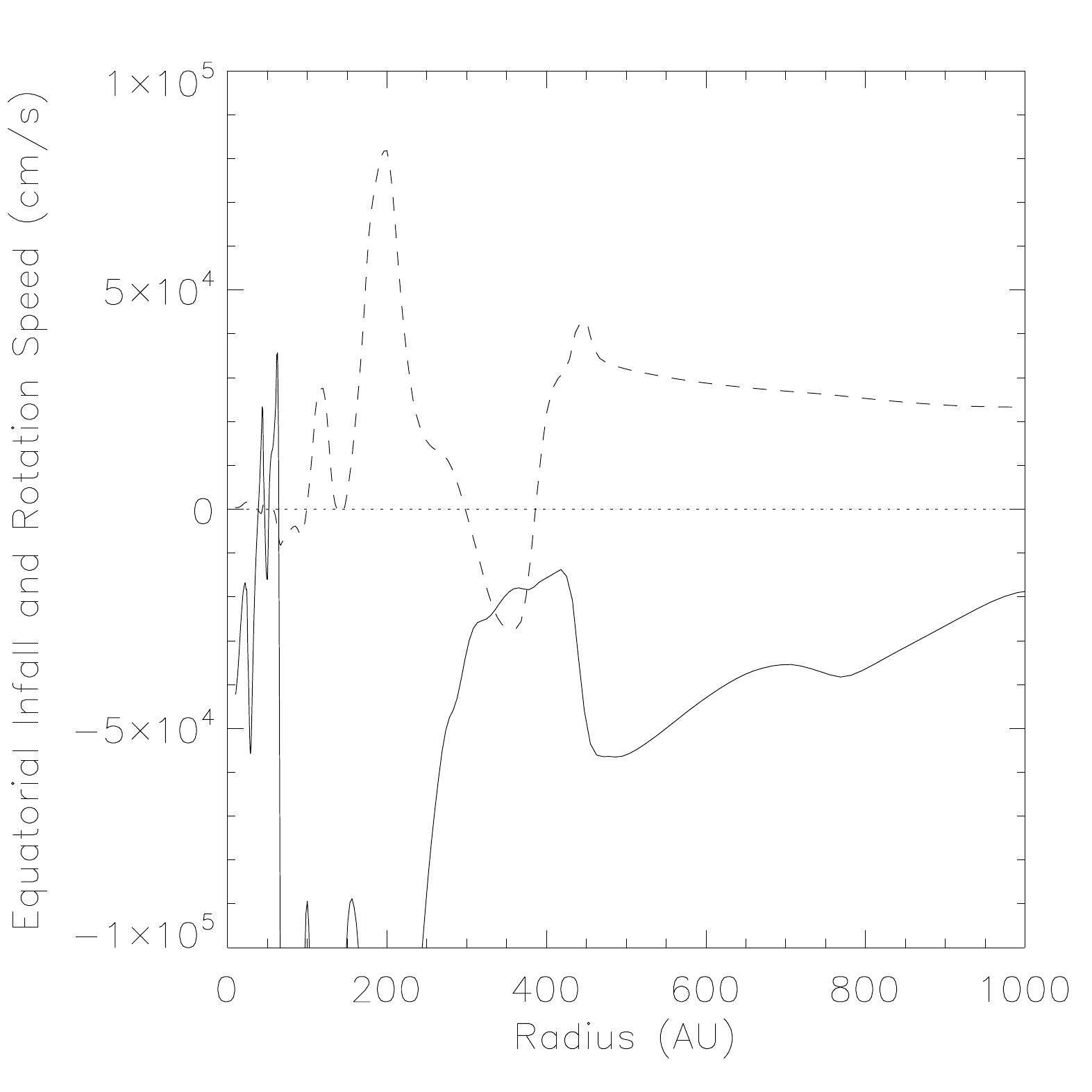}
\caption{Equatorial infall $v_r$ (solid line) and rotation $v_\phi$ (dashed)
for the MHD model with classic resistivity, showing the lack of a
rotationally supported disk. Note the counter-rotation inside
$400\au$, with rotation speed often decreasing (rather than increasing)
toward the origin. The spikes of positive radial velocity at
$\sim 50$ are related to episodic reconnection events.}
\label{classic_vrvphi}
\end{figure}

Figure \ref{classic} displays a snapshot of the simulation
at the representative time $10^{12}\second$. A dense flattened
structure develops in the inner equatorial region. It is,
however, fragmented as in the ideal MHD case, indicating
that the classic resistivity is too small to completely
suppress the numerical reconnection of field lines. The
dense blob at radius $\sim 400\au$, in particular, is
similar to the outermost blob in Figure \ref{IdealMHD}, and
is due to an early reconnection event. The artificial
reconnection events make it difficult to carry out
detailed analysis. Nevertheless, it is easy to show that
the dynamics of the inner equatorial region is far from
that of a rotationally supported disk. First, the equatorial
material inside $\sim 400\au$ is often counter-rotating,
with a rotation speed that sometimes decreases toward the center
(rather than increasing
as would be expected for a Keplerian disk: see Figure \ref{classic_vrvphi}).
Second, the radial
motion in the region is mostly supersonic ($> 2\times
10^4\cm\second^{-1}$). The rapid infall is an indication that the
dense, flattened structure is a magnetically produced
pseudodisk \citep{gs93}, rather than a
rotationally supported Keplerian disk. There is one
region of supersonic {\it outflow} at $\sim 50\au$.
This region, and other outflow events that appeared during the simulation,
are related to episodic reconnection
events, as in the ideal MHD model. In any case,
classic resistivity does not appear capable of weakening
magnetic braking enough to enable the formation
of a rotationally supported disk. Enhanced resistivity
is needed if ohmic dissipation is to save the disk.

\begin{table}
  \caption{Model Characteristics}
  \label{table:models}
\vskip -1.5ex
    \begin{tabular}{p{7.5ex}p{5.5ex}p{3ex}p{3ex}l} \hline \hline
{Model}     & {$B_0${\scriptsize($\mu{\rm G}$)}} & {$\LARGE\,\,\eta$} & {Disk}   & {Notes} \\ \hline
HD  & 0 & no  & yes  & Hydro simulation   \\ \hline
IMHD  & 35.4 & no  & no   & Ideal MHD  \\
CR  & 35.4 & yes  & no   & Classic resistivity given by Eq.\ (\ref{standard}) \\
ER18 & 35.4 & yes  & no   & Enhanced resistivity $\eta=10^{18}$ \\
ER18.5 & 35.4 & yes  & no   & Enhanced resistivity $\eta=3\times 10^{18}$ \\
ER19 & 35.4 & yes  & no   & Enhanced resistivity $\eta=10^{19}$  \\
ER19.5 & 35.4 & yes  & yes   & Enhanced resistivity $\eta=3\times 10^{19}$  \\
ER20 & 35.4 & yes  & yes   & Enhanced resistivity $\eta=10^{20}$ \\ \hline
IMHDw  & 10.6 & no  & no  & Weaker field, ideal MHD\\
ER18w & 10.6 & yes  & no  & Weaker field, $\eta=10^{18}$ \\
ER18.5w & 10.6 & yes  & yes  & Weaker field, $\eta=3\times 10^{18}$ \\
ER19w & 10.6 & yes  & yes   & Weaker field, $\eta=10^{19}$  \\
ER19.5w & 10.6 & yes  & yes   & Weaker field, $\eta=3\times 10^{19}$  \\
ER20w & 10.6 & yes  & yes   & Weaker field, $\eta=10^{20}$ \\ \hline
\multicolumn{5}{p{0.94\columnwidth}}{{\sc Note} --- ``Disk'' in column 4 refers to ``rotationally
supported disk'' at time $t=10^{12}\second$.  Enhanced resistivities
$\eta$ are given in cgs units ($\etaunit\,$).}
  \end{tabular}
\end{table}

\begin{figure}[t!]
\includegraphics[width=\columnwidth]{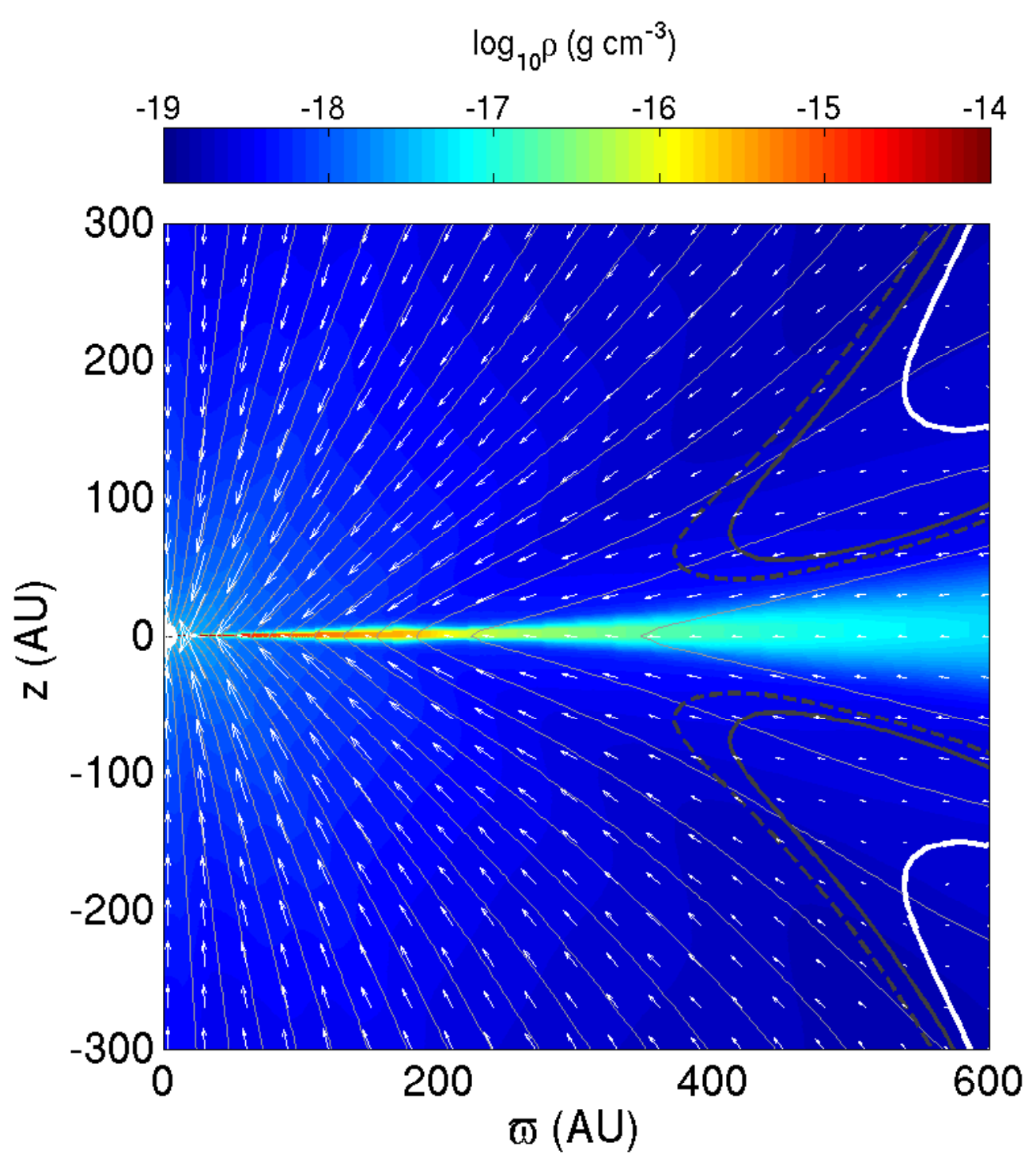}
\caption{Same as Figures \ref{hydro} and \ref{IdealMHD} but for
Model ER18 with an enhanced resistivity $\eta=10^{18}\etaunit$.}
\label{ER18}
\end{figure}

\begin{figure}
\includegraphics[width=\columnwidth]{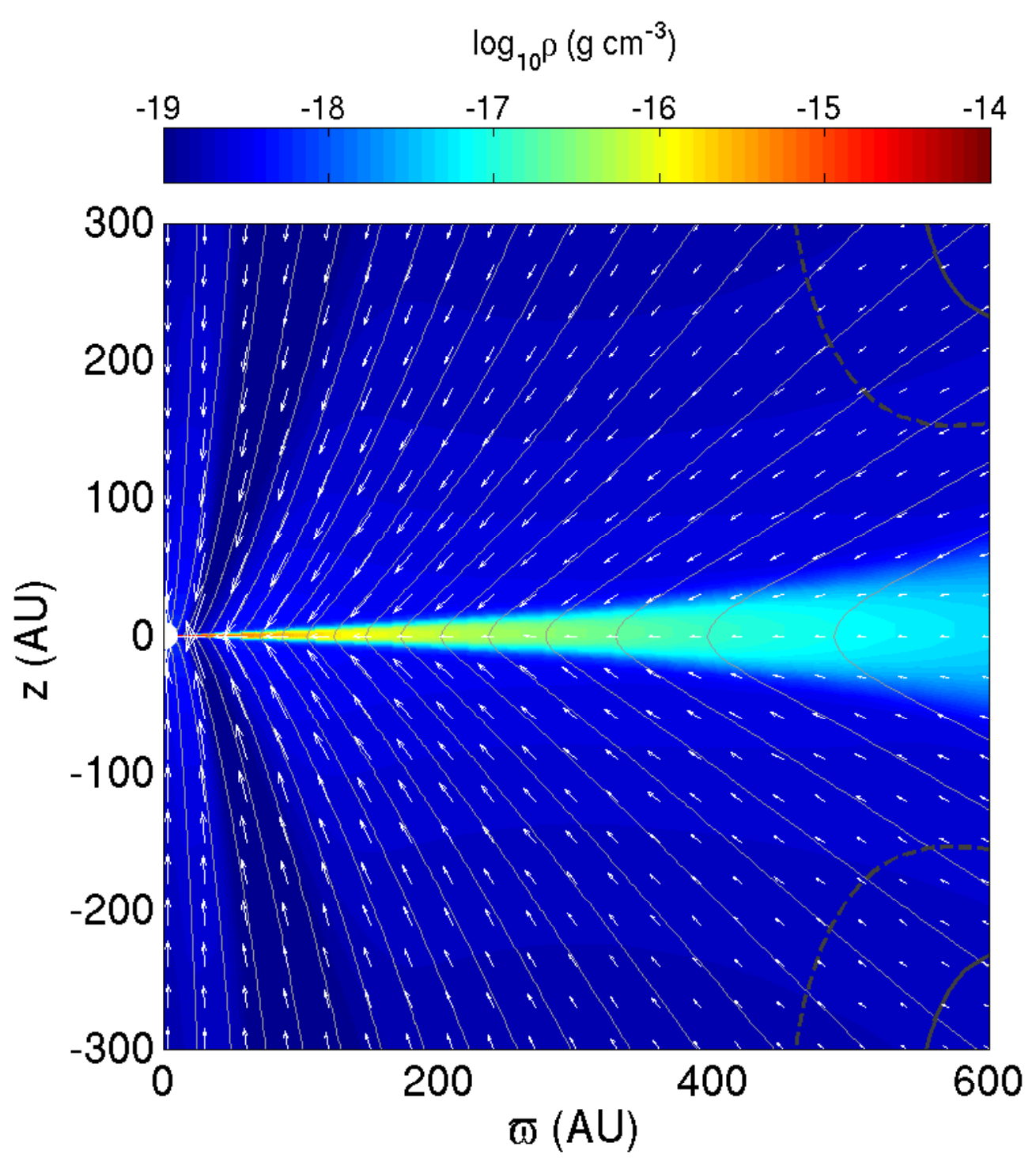}
\caption{Same as Figures \ref{hydro} and \ref{IdealMHD} but for
Model ER19 with an enhanced resistivity $\eta=10^{19}\etaunit$.}
\label{ER19}
\end{figure}

\begin{figure}[t!]
\includegraphics[width=\columnwidth]{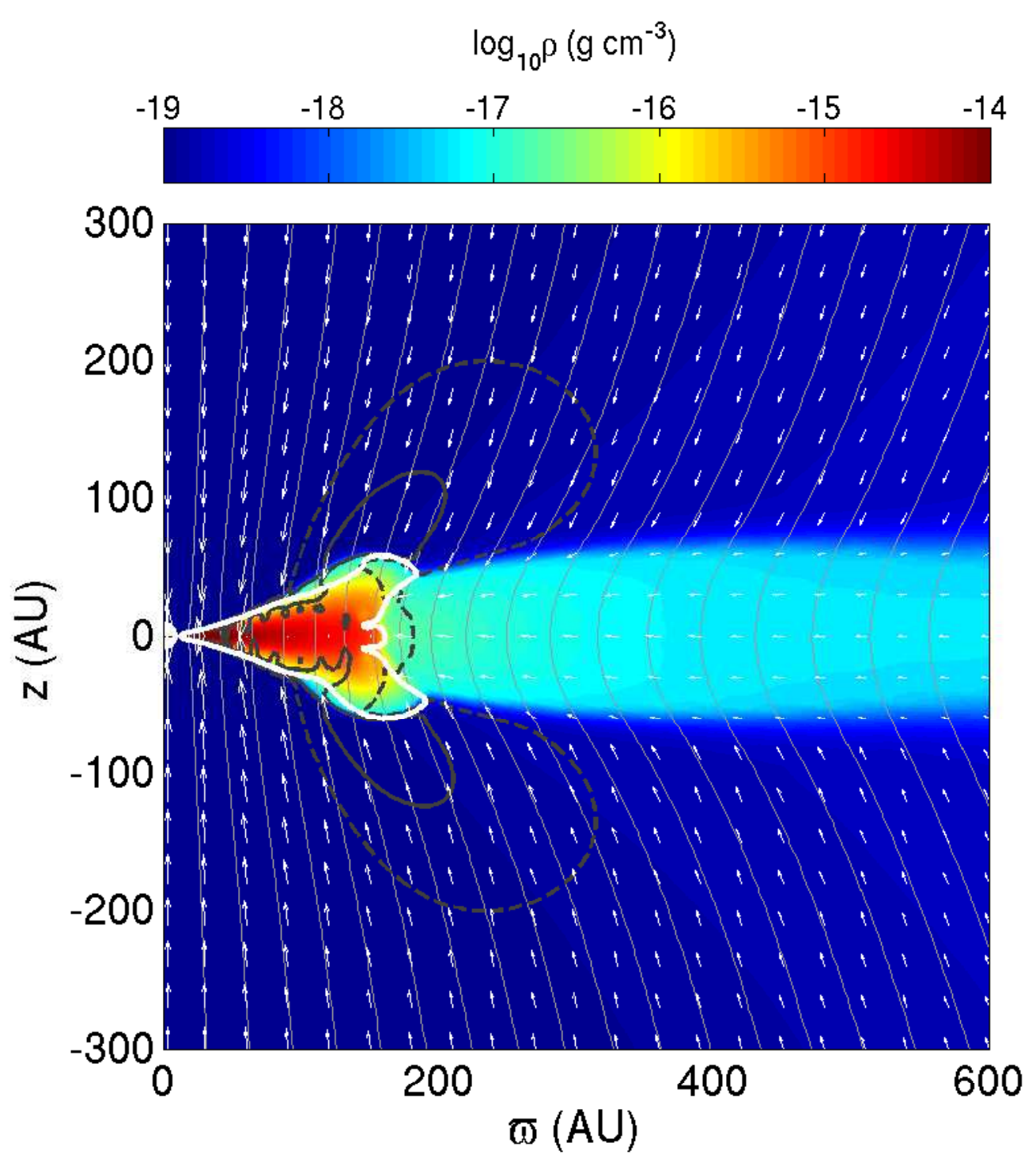}
\caption{Same as Figures \ref{hydro} and \ref{IdealMHD} but for
Model ER20 with an enhanced resistivity $\eta=10^{20}\etaunit$.}
\label{ER20}
\end{figure}

\begin{figure}
\includegraphics[width=\columnwidth]{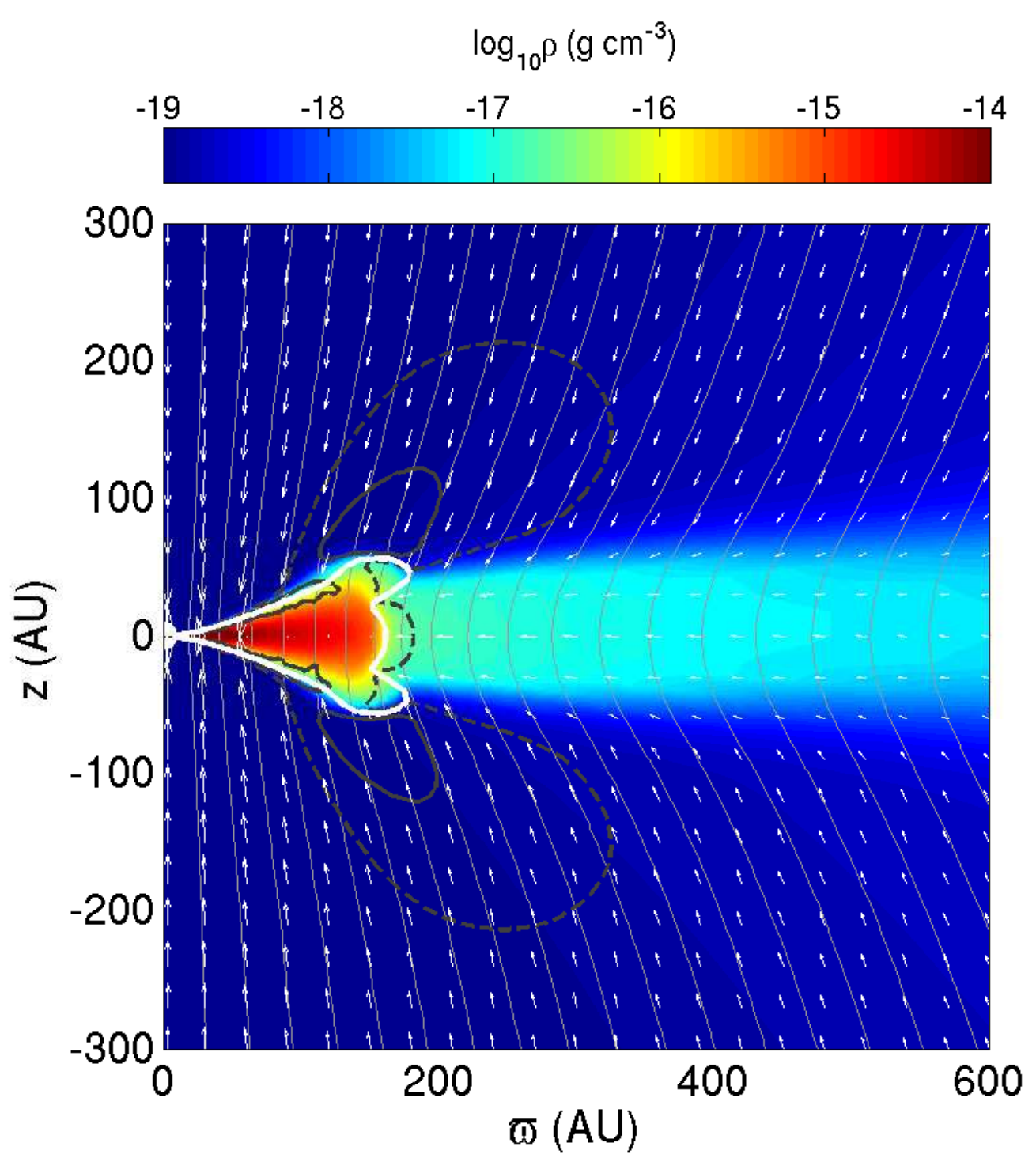}
\caption{Same as Figure \ref{ER20}, with reduced space resolution.}
\label{ER20r}
\end{figure}

\section{Enhanced Resistivity and Disk Formation}
\label{sec:er}
It is well known that anomalous resistivity orders of magnitude
larger than the classic value is required to explain the 11-year
sunspot cycle (\citealp{p07}, \Section{}9.3). Large anomalous resistivity
may also exist in star-forming molecular cloud cores at high
densities \citep{nh85}, although its magnitude
is uncertain. Here we consider the simplest case of a spatially
uniform resistivity of a range of values and for two different
magnetic strength $B_0=35.4$ and $10.6\muG$. The models are
summarized, together with the ones discussed earlier, in Table \ref{table:models}.
We will concentrate on three representative, stronger field
($B_0=35.4\muG$) models with $\eta=10^{18}$ (Model ER18),
$10^{19}$ (ER19), and $10^{20}\etaunit$ (ER20),
respectively. The weaker field models yield qualitatively similar
results, and will be discussed briefly toward the end of the
section.

\begin{figure}[t]
\includegraphics[width=\columnwidth]{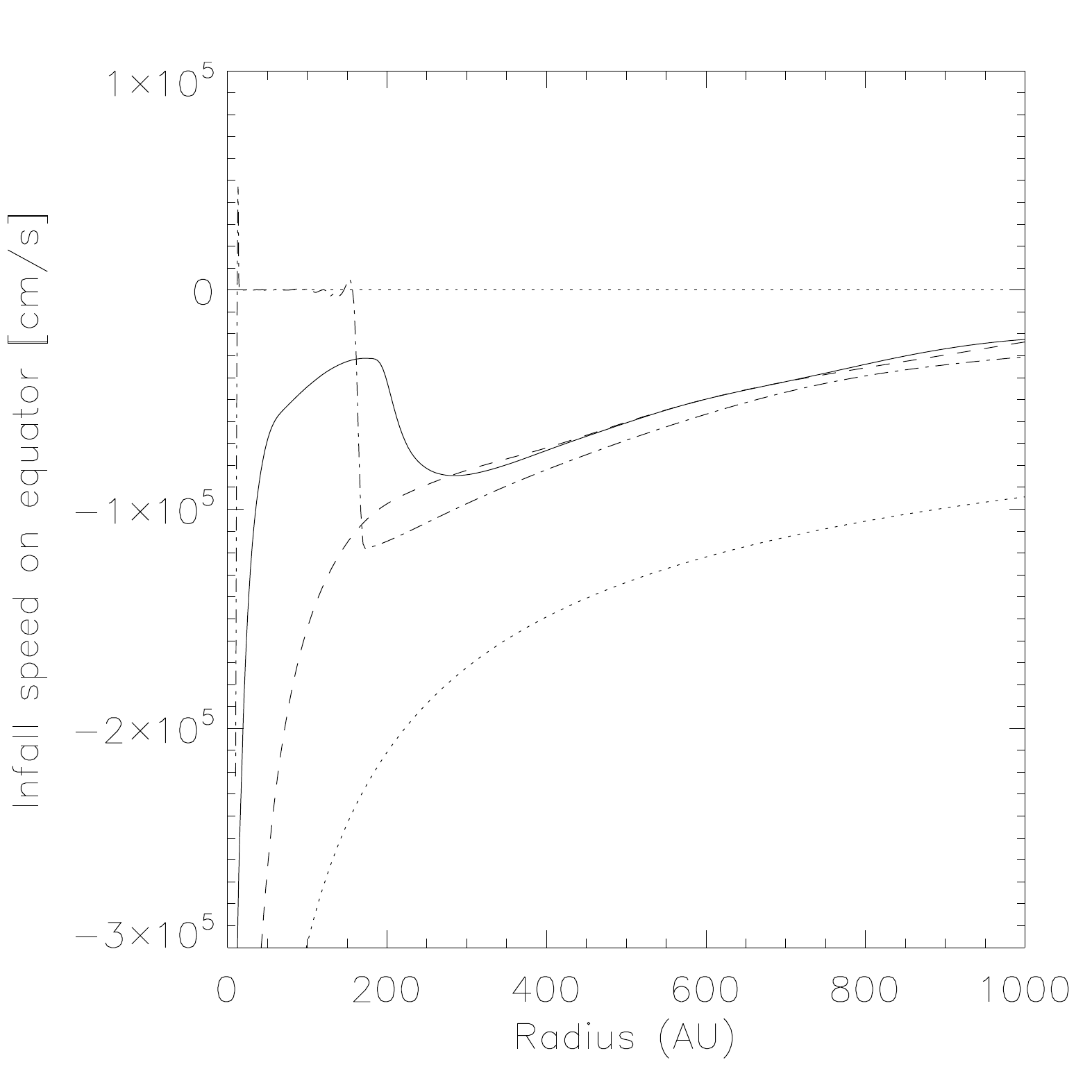}
\caption{Equatorial infall speed for Model ER18
  ($\eta=10^{18}\etaunit$, solid line), ER19
  ($\eta=10^{19}\etaunit$, dashed), and ER20
  ($\eta=10^{20}\etaunit$, dash-dotted line) at the representative
time $t=10^{12}\second$. The free-fall speed is plotted as the lower
dotted line for comparison.}
\label{vr_all}
\end{figure}

\begin{figure}
\includegraphics[width=\columnwidth]{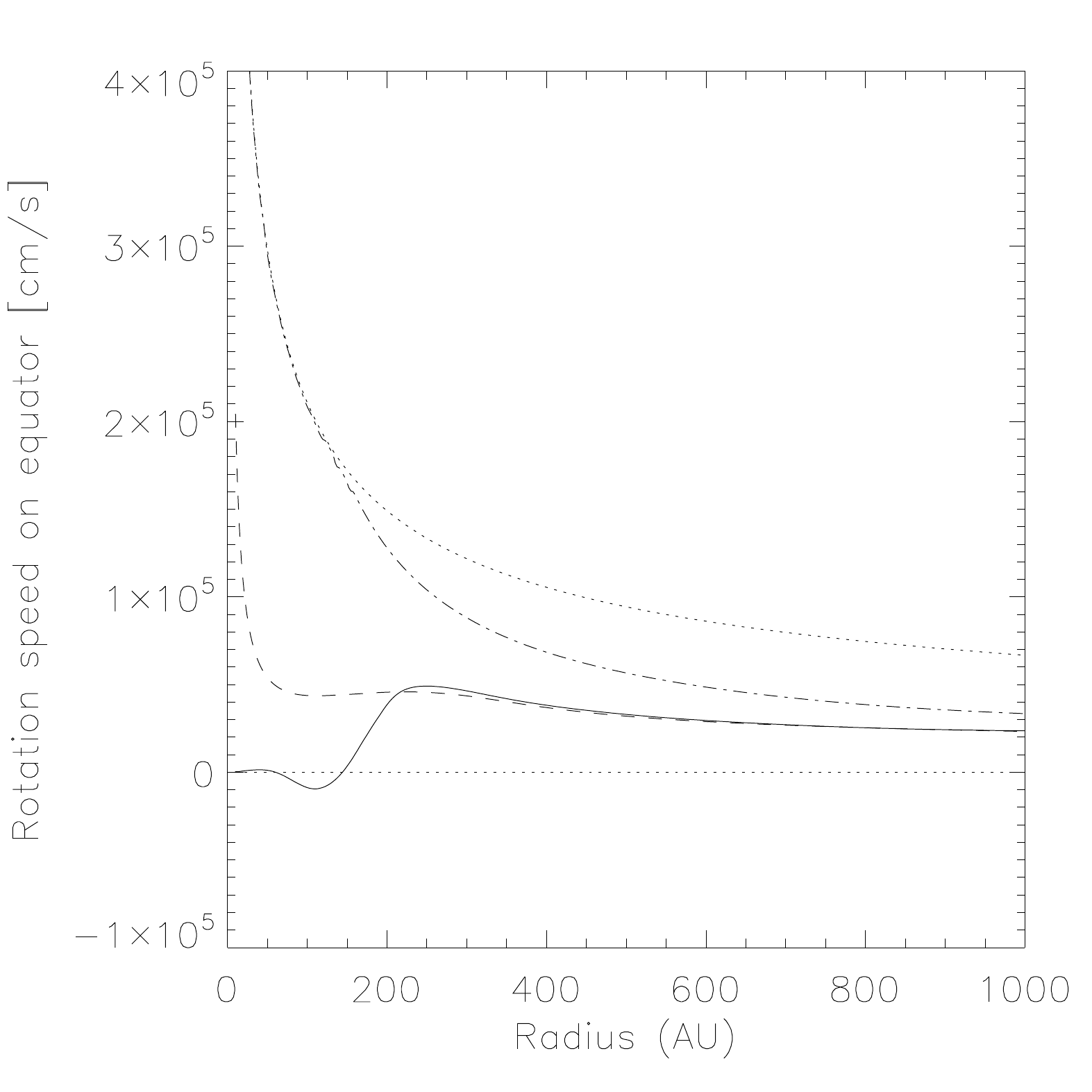}
\caption{Same as Figure \ref{vr_all} but for the equatorial rotation
  speed. The Keplerian speed is plotted as the upper dotted line
for comparison.}
\label{vphi_all}
\end{figure}

\begin{figure}
\includegraphics[width=\columnwidth]{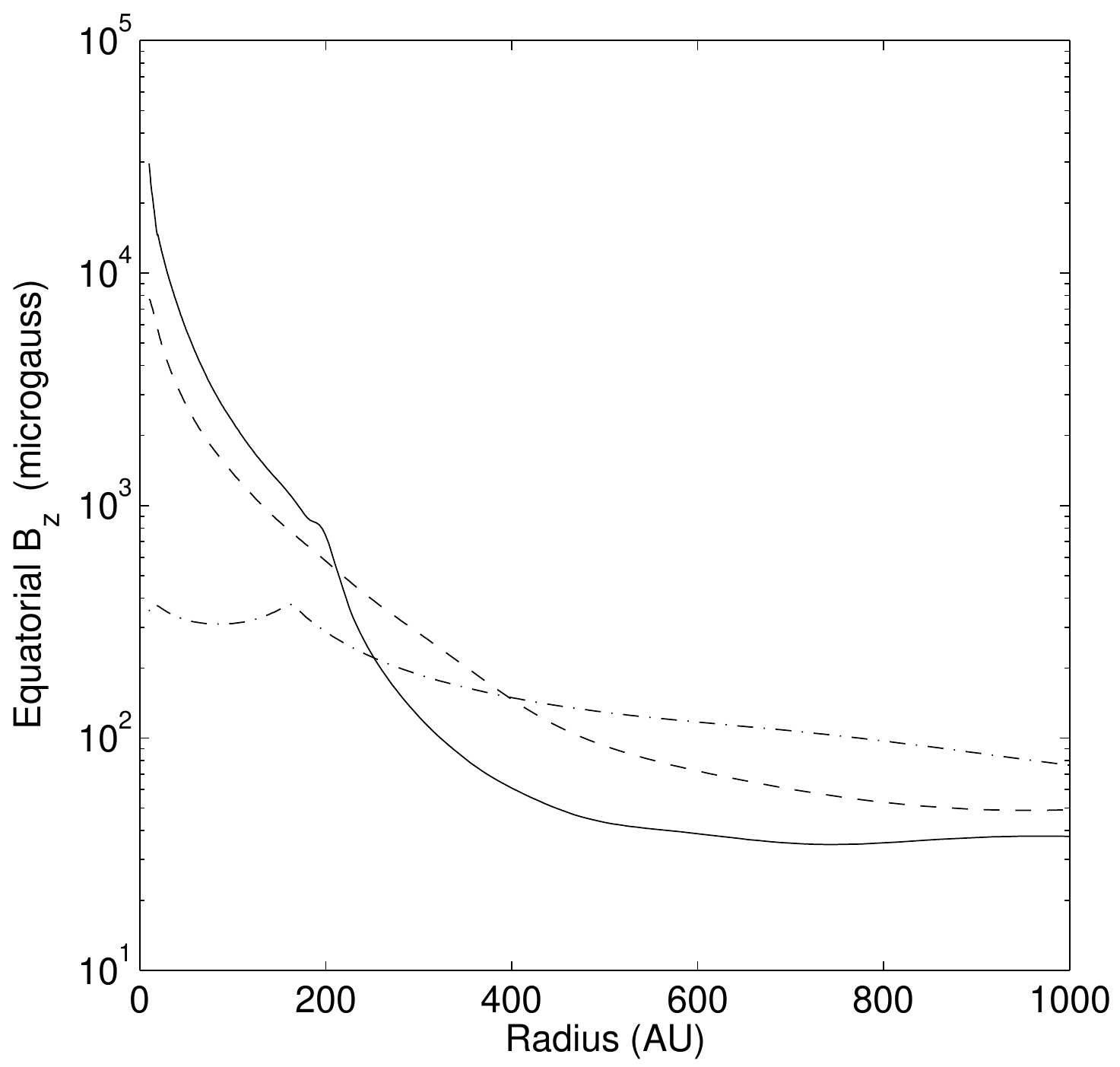}
\caption{Same as Figure \ref{vr_all} but for the vertical
magnetic field component $B_z=-B_\theta$ on the equator.}
\label{Bz_all}
\end{figure}

Figures \ref{ER18}--\ref{ER20} show snapshots of
the three models with enhanced resistivity, and Figures
\ref{vr_all}--\ref{Bz_all} display their infall and rotation speeds and
vertical field strength on the
equator. Additionally, Figure \ref{ER20r} shows the effect on Model ER20
of reducing numerical resolution by a factor of $2$ in both $r$ and $\theta$.
Compared to the models without resistivity (Model HD shown
in Figure \ref{hydro}) and with the classic resistivity (Model CR,
Figure \ref{classic}), the mass distribution and velocity field
are much smoother, indicating that the enhanced resistivity in
these models has largely eliminated the numerical reconnection.
The suppression facilitates quantitative analysis of the
simulations.

The mass distributions in the two lower resistivity models (Models ER18
and ER19) appear similar. Both have a dense equatorial region sandwiched
from above and below by a more diffuse infalling envelope. The infall
in the inner part of the envelope (within a few hundred AUs) appears
to be guided by the field lines. It strikes the surfaces of the dense
equatorial structure at a supersonic speed, forming two ``accretion
shocks.'' These are reminiscent of the classic accretion shocks
around Keplerian disks formed in non-magnetic simulations (e.g.,
\citealp{b90}; see Figure \ref{hydro}). Their production
is due, however, mainly to the magnetic field rather than rotation.
Indeed, the same shocks form even in the absence of any rotation.
For this reason, we will term these shocks ``the magnetically
induced accretion shocks.'' They surround a magnetically induced
``pseudodisk'' \citep{gs93} rather than a rotationally supported Keplerian
disk.

Figure \ref{vr_all} shows that the dense equatorial structure in
Models ER18 and ER19 contains a pseudodisk rather than a rotationally
supported disk. In both cases, the infall is supersonic
($>2\times 10^4\cm\second^{-1}$)  within $10^3\au$ of the central object,
although there is significant difference in the infall speed
within $\sim 200\au$: the equatorial material in the less
resistive model ER18 falls inward much more slowly than that
in the more resistive model ER19. The prominent deceleration
of infalling material around $\sim 200\au$ in Model ER18 is
reminiscent of the ambipolar diffusion-induced C-shock (or
C-transition) first discussed in \citeauthor{lm96} (\citeyear{lm96}; see also
\citealp{ck98}; \citealp{l99}; \citealp{kk02}; \citealp{tm07}
and especially Figure 3 of \citealp{ml09}).
The physical origin is also similar. Just as ambipolar
diffusion, the ohmic dissipation enables the field
lines to diffuse relative to the material that plunges into the
central object. The net effect is the formation of a strong
(poloidal) field region at small radii, where the inward
advection of the field lines by infalling material is balanced
by the resistivity-induced outward magnetic diffusion. A
relatively small resistivity such as that in Model ER18
allows for relatively larger magnetic gradients and Lorentz forces,
which decelerate the infalling material more strongly.
When the resistivity is small, the magnetic
field remains a significant barrier to protostellar accretion.

The magnetic barrier weakens as the resistivity increases.
In Model ER19, which is ten times more resistive than
Model ER18, the ohmic dissipation-induced deceleration has
all but disappeared. This is because the higher resistivity
can support only a smaller current, making it hard to
effectively ``trap'' the magnetic flux that is dragged
to the small radii by the collapsing flow. Nevertheless,
the infall speed is considerably smaller (by about a
factor of two or so) than the free-fall speed, which is
also plotted in Figure \ref{vr_all} for comparison. The
sub-free-fall motion comes about because the equatorial
infall is retarded by a combination of centrifugal and
magnetic forces, although the magnetic forces appear
to be the more important contributor, since the rotation
speed shown in Figure \ref{vphi_all} is about 1/3 of the
Keplerian speed or less (yielding a centrifugal force
an order of magnitude weaker than the gravity).

Note from Figure \ref{vphi_all}, for the least resistive
model ER18,
the equatorial rotation speed within $\sim 200\au$ of
the origin decreases, rather than increases, toward the
center. Obviously, the decrease is due to magnetic braking,
which is apparently so efficient as to produce a region
of counter-rotation at $\sim 100\au$. The counter
rotation is reminiscent of the case of magnetized core
collapse in the presence of ambipolar diffusion with
a relatively high cosmic ray ionization rate
$10^{-16}\second^{-1}$  studied by \citeauthor{ml09} (\citeyear{ml09};
see the bottom panel of their Figure 4). In both cases,
the efficient braking is due to the strong poloidal
field trapped by infall at the small radii. It removes
essentially all of the angular momentum of the
equatorial material that crosses the inner boundary
of the computation domain. In the more resistive model
ER19, the equatorial rotation speed inside $\sim 200\au$
is initially kept nearly constant, before increasing
rapidly. The peak rotation speed remains well below the
local Keplerian speed, indicating that a rotationally
supported disk is not formed. We conclude that the dense,
flattened, equatorial structure produced in both
Models ER18 and ER19 feature dense pseudodisks, but no rotationally
supported dense disks. These pseudodisks are thinner than the
Keplerian disk shown in Figure \ref{hydro} for the hydro case because
of vertical compression by pinched field lines, and
smaller column densities
(due to the fast infall rate inside the pseudodisk), corresponding
to reduced scale height for
a given gas density.  The absence of a rotationally
supported disk in these two cases (which have resistivities
much larger than the classical value) supports our results in
\Section{}2.2 and 3 that disk formation is suppressed by magnetic
braking in the ideal MHD limit and in the presence of the
classical resistivity, even though these results were affected
by numerical reconnection.

The most resistive model ER20 differs drastically from the lower
resistivity cases. It produces a much denser disk of radius
$\sim 150\au$ at the representative time $t=10^{12}\second$
(see Figure \ref{ER20}), which turns out to be rotationally
supported. The strongest evidence for rotational support
comes from Figure \ref{vphi_all}, which shows that the
rotation speed at the equator is nearly identical to
the Keplerian speed inside $\sim 150\au$. The support
is also evident in Figure \ref{vr_all}, which shows that the
infall has nearly stopped at small radii (the spike near
the inner edge of the computation grid is likely related
to the outflow boundary condition, which is not ideal
for a rotationally supported disk; it was present in
the hydro case as well). The thick dashed and solid
dark gray lines in Figure \ref{ER20}, which mark the location
of $|v_\phi|=0.9\,v_K$ and $v_K$, respectively, also make the
case. The disk has a mass $\sim 0.01\msun$ at $t=10^{12}\second$,
corresponding to an average accretion rate (from the envelope
to the disk) $\sim 5\times 10^{-7}\msun\yr^{-1}$.
As in the pure hydro case (\Section{}2.1), little of the disk
material passes through the disk onto the central object
in our axisymmetric simulations. Gravitational torques
may be needed to drive the further evolution of such disks
(in three dimensions).

Another feature of Model ER20 is that the poloidal field
lines pass through the equatorial region with only a
moderate bending. It is of course a consequence of the
large resistivity, which allows the field lines to diffuse
outward easily relative to the infalling material. This
is especially true inside the inner, Keplerian disk, where
the combination of slow radial motion and large resistivity
means that the magnetic field must be nearly vertical and
uniform (see the dot-dashed line in Figure \ref{Bz_all}).
A consequence of the nearly uniform field is that
there is little vertical compression by the magnetic field,
yielding a Keplerian disk that is thicker than the
magnetically compressed pseudodisks in the less
resistive models (see Figures \ref{ER18} and \ref{ER19}).

Outside the Keplerian disk lies a supersonically infalling
pseudodisk. It is linked to the surrounding envelope by
magnetic field lines; they are separated by the magnetically
induced accretion
shocks. Upstream of the shock, the low density envelope
collapses more or less along the field lines, indicating that
flux-freezing is approximately valid, because the
field lines there are relatively straight and there is
little current to be dissipated by the (large) resistivity.
Downstream of the shock, the dense pseudodisk collapses
horizontally, nearly perpendicular to the (poloidal) field
lines. At the relatively late time shown in Figure \ref{ER20},
the field lines inside the pseudodisk are nearly held fixed
in space, with the inward advection due to infalling material
balanced by the outward resistivity-driven diffusion.

The role of resistivity in enabling disk formation
can be seen clearly in Figure \ref{Bz_all}, where the
vertical component of the magnetic field on the equator,
$B_z$, is plotted as a function of time for all three
models with different enhanced resistivities. The equatorial
infall in the pseudodisk tends to compress the poloidal
field lines, leading to a sharp increase in the poloidal
field strength at small radii. Resistivity, on the other
hand, tends to smooth out the field distribution. The net
effect is that the poloidal field strength at small radii
depends strongly on the magnitude of the resistivity,
being larger for smaller resistivities. Since most of the
magnetic braking occurs at relatively small radii, where
the infalling material attempts to spin up the fastest
to conserve angular momentum, the stronger poloidal
fields in the two less resistive cases (models ER18 and ER19)
produce a stronger braking, which is enough to suppress
the formation of a rotationally supported disk. The most
resistive case (model ER20) has the weakest poloidal
field at small radii (where magnetic braking matters
most in the model) which, together with a large slippage between
the field lines and matter in the azimuthal direction
(which makes it hard to generate a toroidal field),
makes the magnetic braking ineffective and disk
formation possible. That the disk in model ER20 forms
as a result of enhanced resistivity rather than
numerical diffusion is supported by the fact that
a nearly identical disk is formed in the lower
resolution run (see Figure \ref{ER20r}).

One can roughly estimate the size of the resistivity-enabled
disk as follows. In order for the formed rotationally
supported disk to survive against magnetic braking, the
disk rotation must not twist up the poloidal magnetic
field that threads the disk too strongly. This is
satisfied when the magnetic diffusion speed becomes
comparable to the Keplerian speed, which leads to
the condition $\eta\sim v_K H$, where $H$ is the
vertical scale length over which the toroidal magnetic
field varies. If we take $H$ to be the disk pressure
scale height (ignoring possible magnetic compression
of the disk), and use the standard result for a
thin-disk $H/r \sim a/v_K$ (where $r$ is the radius
and $a$ is the isothermal sound speed), then $\eta \sim
r a$ or $r \sim \eta/a \sim 3.3\times 10^2\au$
(for $\eta=10^{20}\etaunit$ and $a=2\times
10^4\cms$). This is about a factor of two
larger than the disk radius that we found numerically
in Figure \ref{ER20}. We should caution the reader
that the above estimate did not take into account
the detailed magnetic geometry and braking
efficiency, which depends on the field strength
(see below).
For a weak enough field, a disk can form independent
of the magnitude of the resistivity.

Besides the three representative models discussed above,
we have carried out two additional simulations with
$\eta=3\times 10^{18}$ (Model ER18.5) and
$3\times 10^{19}\etaunit$ (Model ER19.5). As one
would expect based on the results for Models ER18 and
ER19, no rotationally supported disk forms in Model
ER18.5. There is a small ($\sim 40\au$ in radius)
rotationally supported disk in the more resistive Model
ER19.5 at the fiducial time $t=10^{12}\second$. The disk
shrinks with time, however. By $t=2\times 10^{12}\second$,
its radius decreases to $\sim 25\au$; it may disappear
altogether at a later time. The trend indicates that
$\eta=3\times 10^{19}\etaunit$  is probably close
to the critical resistivity $\eta_c$ needed for disk
formation.

The value for the critical resistivity $\eta_c$ depends
on the initial field strength $B_0$. This is to be
expected since, in the limit of infinitely weak field,
rotationally supported disks can form without any
(enhanced) resistivity at all (i.e., $\eta_c=0$). For a
moderately magnetized dense core of $B_0=10.6\muG$,
we find that the transition between the formation of
a rotationally support disk and its suppression
occurs between $\eta=10^{18}$ (Model ER18w) and
$3\times 10^{18}\etaunit$ (Model ER18.5w).
This value of $B_0$ is probably close to the lower
limit to the field strength in dense cores, judging
from the fact that the median field strength for
the more diffuse, cold neutral medium (CNM) of
atomic gas is $\sim 6\muG$ \citep{ht05}
and that the directly measured line-of-sight
component of the magnetic fields in a number of
dense cores is of this order or higher
\citep{ct08}; the full strength of the core
magnetic field is likely significantly higher. It
is therefore reasonable to expect the critical value
$\eta_c$ for disk formation in cloud cores magnetized to
a realistic level to lie somewhat between $\eta_c \sim
3\times 10^{18}$ and $\sim 3\times 10^{19}\etaunit$.
In what follows, we will take $\eta_c=10^{19}\etaunit$
as the characteristic critical value, with the understanding
that it depends somewhat on the field strength, and can be
uncertain by a factor of $\sim 3$ in either direction.

\section{Discussion and Conclusion}

Our most important result is that classical resistivity
alone appears unable to weaken magnetic braking enough
to allow a rotationally supported disk to form for a
realistic level of core magnetization, and that an
enhanced resistivity of order $\eta_c=10^{19}\etaunit$
is needed.

A large, enhanced resistivity (of order
$2\times 10^{20}\etaunit$)
was advocated by \citet{sglc06} to solve
the so-called ``magnetic flux problem'' in star formation.
They introduced the concept of the Ohm sphere, with a radius:
\begin{equation}
r_{\rm Ohm}= \frac{\eta^2}{2G M_*}
\label{ohmsphere}
\end{equation}
where $M_*$ is the central stellar mass. It can be
obtained by setting the characteristic magnetic
diffusion speed $v_{\rm d}=\eta/r$ equal to the
free-fall speed $v_{\rm ff}= (2 G M_*/r)^{1/2}$.
In the limit of a weak magnetic field (so that the magnetic
forces do not affect the accretion dynamics) and zero
rotation, \citet{sglc06} showed that the magnetic field becomes
more or less uniform inside the Ohm sphere; the sphere
may be thus regarded as the region of complete magnetic
decoupling. If a rotationally supported disk were to
form inside the Ohm sphere, it would be protected against
strong braking by the decoupling.

If the Ohm sphere must be larger than the rotationally
supported disk, then to form a disk of $10^2\au$ in
radius around a half-solar mass star, the resistivity
must be $\eta_0 =\sqrt{2G M_* r_d} \sim 4.5 \times
10^{20}\etaunit$.
Our actual, numerically obtained, value of resistivity
for disk formation is one to two orders of magnitude
smaller, depending on the magnetic field strength. The
discrepancy indicates that the
simple considerations leading to the Ohm sphere need
to be modified. One such modification is that the (radial)
magnetic diffusion speed $v_d$ is expected to be larger
than $\eta/r$, because the field lines bend sharply
inside the pseudodisk, with a field variation scale
$\Delta$ much less than $r$ (see Figs.\ \ref{ER18} and
\ref{ER19}). The other is that the infall speed can be
significantly below the free-fall value (see Fig.\ \ref{vr_all}).
Both modifications make it easier for the field lines
to diffuse out, lowering the critical value for disk
formation compared with the simple estimate based on
the concept of Ohm sphere.

Nevertheless, the critical resistivity for disk formation
is still a few orders of magnitude higher than the
classical value. How such an enhanced resistivity may
arise is unclear (see \citealp{p79}). One possibility
is turbulent resistivity, which has been discussed, for
example, in the context of magnetized accretion disks
(e.g., \citealp{lpp94}; \citealp{gg09}). There
are at least two possible origins for turbulence on the
disk formation scale (of $10^2$--$10^3\au$). First, the
dense cores that collapse into stars are observed to
have non-thermal motions on the tens of thousands of AU
scale, which can be interpreted as a subsonic
turbulence of order half the sound speed \citep{bt07}.
It is likely that the turbulent motions would
continue down to smaller scales as the core collapses.
Second, the collapsing flow can be significantly retarded
by magnetic forces (see, e.g., the infall speed for Model
ER18 in Figure \ref{vr_all}). The magnetic support of
the collapsing material against gravity may lead to
interchange instability in 3D, which could develop
into a turbulence, and might itself lead to
an effective turbulent resistivity.  Whether turbulent resistivity can
enable the formation of rotationally supported disks
or not requires high-resolution 3D simulations,
perhaps along the line of \citet{klvo09}. Another
possibility is anomalous resistivity associated with,
for example, current-driven instabilities \citep{nh85}.
Whether these and other enhancements
of resistivity can enable disk formation remains to
be determined. If not, other means of enabling disk
formation, such as the Hall effect and outflow
stripping of the protostellar envelope, need to be
explored.

Although our idealized calculations indicate that a
considerable enhancement in resistivity is needed
if ohmic dissipation alone is to enable the formation of
rotationally supported disks during the protostellar
accretion phase of low-mass star formation, the exact
value for the required enhancement is uncertain, and
can be better determined with future refinements. One
refinement is to include the self-gravity, which would
also allow us to follow the formation and evolution
of the dense magnetized core to the point of point-mass
formation. The prestellar core evolution would provide
a more self-consistent initial condition for the disk
formation calculation during the protostellar accretion
phase. An integrated treatment of both core formation
and protostellar accretion may enable us to determine
the distribution of the matter surrounding the dense
core, which can aid the core material in removing
the angular momentum from the forming disk, especially
if the core is highly flattened along the field lines
(\citealp{bm94}; \citealp{kk02}; see e.g., \citealp{t09} for statistical
evidence of core flattening). Another refinement,
which we plan to do, is to study the combined effect
of resistivity and ambipolar diffusion on disk
formation.
In addition, it would be desirable to find a better way to
quantify the numerical reconnection, originated physically
from the strong collapse-induced magnetic pinching, that
affects the ideal MHD and low resistivity simulations.
We conclude that how protostellar disks
form around accreting protostars remains a mystery.

\acknowledgments

We acknowledge support by the Theoretical
Institute for Advanced Research in Astrophysics (TIARA),
and by the National Science Council of Taiwan through
grant NSC97-2112-M-001-018-MY3.
This work is supported in part by NASA ATP (NNG06GJ33G)
and Origins grants. We acknowledge useful conversations
with Roger Blandford, Arieh K\"onigl, Chris McKee, Jon McKinney,
Nir Shaviv, and Ron Taam.

\end{document}